\documentclass[a4paper,11pt]{article}
\usepackage{physics}
\pdfoutput=1 

\usepackage{jcappub} 

\usepackage[T1]{fontenc} 

\usepackage{amssymb}
\usepackage{amsmath}    
\usepackage{graphicx}   
\usepackage{verbatim}   
\usepackage{color}      
\usepackage{subfigure}  
\usepackage{hyperref}   
\usepackage{float}     
\usepackage{mdwlist}

\newcommand\be{\begin{equation}}
\newcommand\ba{\begin{eqnarray}}
\newcommand\ee{\end{equation}}
\newcommand\ea{\end{eqnarray}}

\title{String gases and the swampland}

\author{Samuel Laliberte and Robert Brandenberger}
\affiliation{Physics Department, McGill University, 3600 University Street, Montreal, QC, H3A 2T8, Canada}

\emailAdd{samuel.laliberte@mail.mcgill.ca}
\emailAdd{rhb@physics.mcgill.ca}

\abstract{In this paper, we study some aspects of moduli stabilization using string gases in the context of the Swampland.  In the framework which we derive, the matter Lagrangian for string gases yields a potential for the size moduli which satisfies the de Sitter conjecture with the condition $\frac{|\nabla V|}{V} \geq \frac{1}{\sqrt{p}}\frac{1}{M_p}$, where $p$ is the number of compactified dimensions.  Moreover, the moduli find themselves stabilized at the self-dual radius, and gravity naturally emerges as the weakest force.}

\begin{document} 
\maketitle
\setlength{\parindent}{1em}
\setlength{\parskip}{1em}


\section{Introduction}

There has recently been a lot of discussion concerning what types of effective field theories for gravity coupled to matter can be consistent with superstring theory. Those that do belong to the {\it landscape}, while those that do not are said to be in the {\it swampland}. Specifically, \cite{Obied:2018sgi, Ooguri:2018wrx} (see also \cite{Krishnan}) suggested that all scalar fields arising from string theory must have a potential which satisfies
\be
\frac{|\nabla V|}{V} \geq \frac{c}{M_p} \; ,
\ee
or
\be
\frac{\min (\nabla_i \nabla_j V)}{V} \leq  - \frac{c'}{M_p} \; ,
\ee
where $c,c'$ are universal constants of order 1 and $M_p$ is the Planck mass (see e.g. \cite{Brennan, Palti} for recent reviews).  This bound, called the {\it de Sitter conjecture}, excludes stable and meta-stable de Sitter vacua in string theory and is motivated by two key arguments.  First, the conjecture is supported by multiple examples in string theory \cite{Obied:2018sgi}.  Second, we expect that when the de Sitter conjecture is violated, a massive tower of states becomes massless leading to a breakdown of the effective field theory.  This second argument was first demonstrated in \cite{Ooguri:2018wrx} using Bousso's \cite{Bousso} covariant entropy bound in an accelerating universe.  It was shown that for field values larger than the Planck scale, an exponentially large number of string states become massless (see also \cite{Ooguri}), saturate the entropy bound and force the scalar field to satisfy the de Sitter criteria.

The de Sitter conjecture is known to severely constrain inflationary scenarios. Hence, it becomes relevant to ask if other cosmological scenarios such as { \it matter bounce} \cite{Bounce} or { \it string gas cosmology} \cite{BV} satisfy the Swampland conjectures.  In the next sections, we will be interested in the case of string gas cosmology, and more particularly moduli stabilization using string gases.

Our work is structured as follows.  In section \ref{sec:modsg}, we introduce a framework for string gases in the context of effective field theories and show how string gases give a potential that stabilizes extra dimensions at their self-dual radius.  In section \ref{sec:sg+grav}, we show how moduli stabilization by string gases implies that gravity must be the weakest force at late times in the universe.  Finally, in section \ref{sec:sg+ds}, we show how the string gas potential satisfies the Swampland de Sitter conjecture.

\section{Moduli sabilisation using string gases}
\label{sec:modsg}

We start with the action of a closed string in a D-dimensional space-time where d dimensions are non-compact and p are compact.  In the "adiabatic" approximation, the Nambu-Goto action can be approximated as the energy of a particle with a mass corresponding to the oscillations of the closed string.  The reader is referred to \cite{Patil:2004zp} \cite{Patil:2005fi} (see also \cite{Watson}) for more information on the adiabatic approximation.  A large number of closed strings with the same mass and momentum give a string gas whose action is given by \cite{Patil:2005fi}
\be
S = - \int dx^D \sqrt{-G} n_{D-1} \sqrt{\vec{p}^2 + M_{\vec{n},\vec{w},N}^2} \; \; ,
\label{eq:sg1}
\ee
where $\vec{p}$ is the momentum of the center of mass of the string in the non-compact dimensions, $n_{D-1}$ is the number density of strings in the $D-1$ spatial dimensions and $M_{\vec{n}, \vec{w}, \vec{N}}$ is the mass of the string, which depends on quantum numbers $\vec{n}$,$\vec{w}$ and $\vec{N}$.  The quantum numbers are, respectively, related to the momentum modes, the winding modes, and the oscillatory modes.  The mass spectrum of the string depends on the choice of compactification.  The simplest example is to consider a torus with a metric $\gamma_{ab}$ which only has diagonal elements.  In this case, the string mass reads
\be
M_{\vec{n},\vec{n},N}^2 = \frac{1}{R^2}\gamma^{ab}n_{a}n_{b} + \frac{R^2}{\alpha'}\gamma_{ab}w^aw^b + \frac{2}{\alpha'}\left(2N + n^aw_a -2 \right) \, .
\label{eq:mass}
\ee
The variable $R$ in the expression for the mass describes the coordinate interval for each cycle of the torus.  In other words, coordinates $x^d = \theta^dR$ parametrise a circle with $0 \leq \theta^d \leq 2\pi$.  We will use this compactification example repeatedly throughout the paper.  For simplicity, we will also set R to unity, with the physical length of the cycles hence being described by the metric $\gamma_{ab}$.  We now wish to find an action that describes a string gas with all possible momentum states and all possible masses in the theory.  We start with a string gas made of one string by choosing the number density
\be
n_{D-1} = \frac{\delta^{D-1}(x^{\tau} - x^{\tau}_{cm})}{\sqrt{G_s}} \; ,
\ee
where $x_{cm}^{\tau}$ is the center of mass of the particle and $G_s$ is the determinant of the spatial part of the metric.  We will work in a regime where $M_{\vec{n},\vec{w},N}$ does not depend on the coordinates of the compactified dimensions.  Therefore, we can integrate over the $p$ internal dimensions to obtain
\be
S = - \int dx^d \sqrt{-g_{00}} \sqrt{\vec{p}^2 + M_{\vec{n},\vec{w},N}^2}\delta^{d-1}(x^{\tau} - x^{\tau}_{cm}) \; .
\label{eq:sg2}
\ee
Here, $g_{00}$ is the zeroth component of the $d$-dimensional metric.  To obtain the desired string gas, we sum over a large number of one string "string gas" actions and overall possible massive states to obtain
\be
S = - \int dx^d \sqrt{-g_{00}} \sum_{\vec{n},\vec{m},N} \sum_{i} \sqrt{\vec{p_i}^2 + M_{\vec{n},\vec{w},N}^2}\delta^{d-1}(x^{\tau} - (x^{\tau}_{cm})_i) \; .
\label{eq:sg3}
\ee
In the thermodynamic limit, the sum in the action above can be viewed as the thermally averaged energy of a relativistic gas with particles of massive states given by $M_{\vec{n},\vec{w},N}$.  The string gas action then reads
\be
S_{sg} = \int dx^d \sqrt{-g} n_{d-1} \langle E_1 \rangle \; ,
\ee
where $\langle E_1 \rangle$ is the thermal average of the energy of a single string and $n_{d-1}$ is the number density of strings in the $d-1$ non-compact spatial dimensions.  The spatial determinant of the metric $\sqrt{g_s}$ is factored in the action by defining $n_{d-1} = (n_{d-1})_0/\sqrt{g_s}$ where $(n_{d-1})_0$ is the comoving number density.  The thermal average $\langle E_1 \rangle$ is computed as 
\be
\langle E_1 \rangle = - \frac{\partial}{\partial \beta} \ln Z_1 , 
\ee
where $\beta$ is the inverse temperature, and $Z_1$ is the finite temperature partition function of a relativistic string with mulitiple massive states given by $M_{\vec{n},\vec{n},N}$:
\be
Z_1 = \sum_{\vec{n},\vec{m},N} V^d \int \frac{d^{d-1}\vec{p}}{(2\pi)^{d-1}}e^{-\beta \sqrt{p^2 + M_{\vec{n},\vec{m},N}^2}} \; .
\ee
Note that we do not include in the sum the tachyonic states of our theory since we expect them to vanish in a supersymmetric generalization of our model.  

For our analysis, we will rewrite the partition function as a sum of Kelvin functions $K_a$ \cite{AMS}.  The result reads
\be
	Z_1 = V^d  \left(2\beta\right)^{-(d-2)/2} \sum_{\vec{n},\vec{m},N} \left(\frac{M_{\vec{n},\vec{m},N}}{\pi}\right)^{d/2} K_{d/2}(\beta M_{\vec{n},\vec{m},N}) \; .
	\label{eq:Z_1.1}
\ee
To account for all the massive states in the theory, we should sum over all the quantum numbers in the equation above.  However, if $\beta \gg M_s^{-1}$, only the states which are massless to first order at the self-dual radius contribute significantly, and the other ones are windowed out by the Boltzman factors.  This "massless modes" approximation was introduced in \cite{Patil:2005fi}, and can be used to significantly simplify the partition function.  Using equation \ref{eq:mass}, we can show that the modes which are massless at the self-dual radius $\gamma_{ab} = \alpha' \delta_{ab}$ obey
\begin{align}
(n^a + w^a)(n_a + w_a) & = 4(N-1) \\
N + w^a n_a & \geq 0 \, .
\end{align}
Among the states which satisfy the condition above, only those which have $N = 1$ and $n^a = - w_a = \pm 1$ are massless to first order.  All modes which are massless to first order are degenerate, so the sum in equation (\ref*{eq:Z_1.1}) is easily evaluated and yields
\be
Z_1 = \tilde{N}\left(2\beta\right)^{-(d-2)/2}\left(\frac{M_{1,-1,1}}{\pi}\right)^{d/2} K_{d/2}(\beta M_{1,-1,1}) \; .
\label{eq:Z_1.2}
\ee
Here, $\tilde{N}$ is the number of states which are massless to first order.  Using the partition function above, the thermal average $\langle E_1 \rangle$ is easily evaluated and yields
\be
\langle E_1 \rangle = M_{1,-1,1} \frac{K_{(d-2)/2}(\beta M_{1,-1,1})}{K_{d/2}(\beta M_{1,-1,1})} + \frac{d-1}{\beta}  \; .
\ee 
So far, we derived an expression for the string energy which holds when the temperature of the universe is below the string mass ($\beta \gg M_s^{-1}$).  This condition provides an operational definition for 'late times' in the universe.  For simplicity, we will take the limit where the temperature goes to zero ($\beta \rightarrow \infty$) and only consider very late times in the universe.  In this limit, $\beta^{-1}$ is very small and the last term can be neglected, and we are left with a term proportional to the string mass.  It is worth noting that the system reaches the late time limit exponentially fast, and that $\beta \leq 10 M_s^{-1}$ is enough to suppress most modes which are not massive to first order. We warn the reader to be careful with the high-temperature limit since the massless mode approximation may not apply in this case, or only apply to certain regions of parameter space.

The expression for the string energy can be simplified further by approximating the Kelvin function in the appropriate limit:
\begin{align}
K_n(\beta M_{1,-1,1}) & \approx \left(\frac{\pi}{2\beta M_{1,-1,1}}\right)^{1/2} e^{-\beta M_{1,-1,1}} & \text{if} \; \beta M_{1,-1,1} \gg 1 \, .
\end{align}
Using the expression above, the thermal energy of the string is easily expressed as the string mass
\begin{align}
\langle E_1 \rangle & \approx M_{1,-1,1} \, .
\end{align}
The result above was previously used to show how size and shape moduli stabilize in the presence of string gases \cite{Brandenberger:2005bd}.  In the latter case, fluxes are required to be present. Making use of a gaugino condensation mechanism it can then be shown that also the dilaton can be stabilized \cite{Danos}, without destabilizing the size and shape moduli.  In our case, expressing the string gas Langragian as the energy/mass density of the strings will prove useful to show why the extra dimension must stabilize at the self-dual radius, and also satisfy the de Sitter conjecture.

\subsection{Moduli stabilisation on the torus}
\label{sec:mod}

It is known that string gases can stabilize the size of extra dimensions \cite{Patil:2004zp, Patil:2005fi, Watson} (see also \cite{Brandenberger:2005fb} for a review), even in the absence of fluxes. In this section, we will show how the size of extra dimensions is stabilized when string gases are included as matter in effective field theories.  For simplicity, we will start with toroidal compactifications, then try to extend our results to more general compactifications.

We start with a simple dimensional reduction ansatz for the metric of a space-time topology $\mathbb{R}^d \times T^p$, were $T^p$ denotes a p-dimensional torus.  The metric reads
\be
ds^2 = G_{MN}dx^{M}dx^{N} = g_{\mu\nu}(x)dx^{\mu}dx^{\nu} + \gamma_{ab}(x)dy^ady^b \; ,
\ee
where $g_{\mu\nu}$ is the metric of the non-compact dimensions, and $\gamma_{ab}$ in the metric of the torus.  We will enforce that $\gamma_{ab}$ only has diagonal elements for the mass spectrum of the string to be given by (\ref{eq:mass}).  With our ansatz, the dimensionally reduced low energy action of bosonic string theory with a string gas reads
\be
S = \frac{1}{2\kappa_0^2}\int d^d X \sqrt{-g}e^{-2\Phi_d}\left[R^d + 4\partial_{\mu}\Phi_d\partial^{\mu}\Phi_d - \frac{1}{4}\partial_{\mu}\gamma_{ac}\partial^{\mu}\gamma^{ab} - 2\kappa^2e^{2\Phi_d}n\langle E_1 \rangle + ...\right] \; .
\ee
Here, we will ignore the $B_{\mu\nu}$ fluxes, which play no role in our analysis.  We will also assume that the dilaton $\Phi_d$ is stabilized and that we can consider it to be a constant.  

The metric $\gamma_{ab}$ will generally involve various scalar fields $\phi^I$ called moduli fields.  The kinetic term of the moduli fields reads
\be
-\frac{1}{4}\partial_{\mu}\gamma_{ac}\partial^{\mu}\gamma^{ab} = - g_{IJ}\partial_{\mu}\phi^I\partial_{\mu}\phi^J \; .
\ee
Therefore, a good parametrization of the moduli fields on the torus is $\gamma_{ab} = e^{2\phi^a}\delta_{ab}$.  With this parametrization, the kinetic term of the moduli fields is Euclidean and the low energy action reads
\be
S = \frac{1}{2\kappa_0^2}\int d^d X \sqrt{-g}e^{-2\Phi_d}\left[R^d - \sum_{a}\partial_{\mu}\phi^a\partial^{\mu}\phi^a - 2\kappa_0^2e^{2\Phi_d}n\langle E_1(\phi^a) \rangle + ...\right] \; .
\ee
It's worth noting that the string gas Lagrangian depends explicitly on the metric $\gamma_{ab}$, and therefore on the moduli fields $\phi^a$.  Thus, the string gas lagrangian yields a potential for the moduli fields given by
\be
V(\phi^a) = 2\kappa_0^2e^{2\Phi_d}n\langle E_1(\phi^a) \rangle \; .
\ee

The moduli will naturally stabilize at the minimum of this potential. To find this minimum, we study the thermal average $\langle E_1 \rangle$ in the late time limit derived in section \ref{sec:modsg}.  We obtain
\be
V(\phi^a) \approx 2\kappa_0^2e^{2\Phi_d}nM_{1,-1,1} \;
\ee
with $M_{1,-1,1}$ satisfying
\be
M_{1,-1,1}^2 = \sum_a e^{-2\phi^a} + \sum_a \frac{1}{\alpha'^2}e^{2\phi^a} - \frac{2p}{\alpha'} \, .
\ee
The expression for $M_{1,-1,1}^2$ is obtained using equation (\ref{eq:mass}) with $\gamma_{ab} = e^{2\phi^a}\delta_{ab}$, $N = 1$ and $n^a = - w_a = \pm 1$.  Notice that the masses of the momentum modes become exponentially large as $\phi^a \rightarrow - \infty$ and that the winding modes become exponentially heavy as $\phi^a \rightarrow  \infty$.  As a result, the moduli $\phi^a$ will be driven towards the self-dual radius of the torus where
\be
e^{2\phi^a} = \alpha' .
\ee
Therefore, all moduli will stabilize at the self-dual point $\phi^a = 1/2\ln\alpha'$ where the gradient of the potential vanishes  ($\partial_{a} V = 0$).

\subsection{Moduli stabilisation on other compact manifolds}

So far, we went over a simple example of how moduli stabilize in toroidal compactifications.  This stabilization process can be conceptually explained for arbitrary compactifications. The key aspect of the process is that the moduli potential $V(\phi^I)$ is directly proportional to the string mass $M$, and that the string mass is composed of a momentum mode component and a winding modes component.  When the size of the extra dimensions shrinks, the mass of the momentum mode starts to dominate and becomes exponentially massive below the self-dual radius.  As a result, the moduli potential increases.  The potential also increases when the size of the extra dimensions increases above the self-dual radius and the winding mode becomes exponentially massive.  As a result, we find that the moduli potential is always minimized at the self-dual radius.  This behavior of the momentum and winding modes is a direct consequence of T-duality, which states that the mass spectrum of string theory is invariant under
\be
R \leftrightarrow \frac{\alpha'}{R} \qquad \qquad n \leftrightarrow m \; .
\ee
Therefore, we expect that the size of extra dimensions will always stabilize at the self-dual radius regardless of the choice of compactification, as long as the mass spectrum of string states obeys the T-duality symmetry.

\section{String gases and gravity as the weakest force}
\label{sec:sg+grav}

We saw that string gases can stabilize the size of extra dimensions to their self-dual radius.  We expect Standard Model physics to emerge when the moduli are stabilized at the minimum of their potential.  As a result, gravity must be the weakest force.  We will show that for string gases, this is always the case to low order in $\alpha'$.  We start with a simple Kaluza-Klein dimensional reduction ansatz:
\be
ds^2 = G_{MN}x^{M}x^{N} = g_{\mu\nu}(x)dx^{\mu}dx^{\nu} + \gamma_{ab}(x)(dy^a + A^a_{\mu}(x)dy^{\mu})(dy^b + A^b_{\nu}(x)dy^{\nu}) \; .
\ee
For simplicity, we will continue to assume $\gamma_{ab}$ only has diagonal non-zero elements.  The low energy action of bosonic string theory in the presence of string gases then reads
\be
S = \frac{1}{2\kappa_0^2}\int d^d X \sqrt{-g}e^{-2\Phi_d}\left[R^d - \frac{1}{4}\partial_{\mu}\gamma_{ab}\partial^{\mu}\gamma^{ab} + 2\kappa_0^2e^{2\Phi_d}n\langle E_1 \rangle - \frac{1}{4}\gamma_{ab}F^a_{\mu\nu}F^{b\mu\nu} + ...\right] \; .
\ee
Again, we will assume $\Phi_d$ to be constant for simplicity.  If we choose $\gamma_{ab} = e^{2\phi^a}\delta_{ab}$, then we saw earlier that the moduli $\phi^a$ stabilize at the self-dual dimensions of the torus where
\be
e^{2\phi^a} =  \alpha' \; .
\ee
In the self-dual configuration, the gravitational coupling and the gauge couplings are respectively given by
\be
\kappa^2 = \kappa_0^2 e^{2\Phi_d} \qquad \text{and} \qquad  (g_A^a)^2 = \frac{2}{\alpha'}\kappa^2 \; .
\ee
Therefore, in the limit where $\alpha' \ll 1$ (the limit in which string perturbation theory is trustworthy), the electromagnetic force will naturally be stronger than gravity, in agreement with the {\it weak gravity conjecture} \cite{WGC}.

\section{String gases and the de Sitter conjecture}
\label{sec:sg+ds}

To see how the de Sitter conjecture is satisfied by the string gas, let us use the ansatz used in Section \ref{sec:mod} for moduli stabilization.  We will also work with $d = 4$ where the string coupling can be written as $\kappa^2 = M_p^{-2}$.  The dimensionally reduced low energy action of bosonic string theory with a string gas then reads
\be
S = \int d^d X \sqrt{-g}\left[\frac{M_p^2}{2}R^d - \sum_{a}\frac{M_p^2}{2}\partial_{\mu}\phi^a\partial^{\mu}\phi^a - n\langle E_1(\phi^a) \rangle + ...\right] \; .
\label{eq:S4d}
\ee
Notice that so far, our definition of the moduli fields $\phi^a$ has made them dimensionless. However, in 4 dimensions, scalar fields have mass dimension 1.  We will restore the right units to the moduli fields by defining the scalar fields $\tilde{\phi}^a = M_p\phi^a$.  These scalar fields have the right mass dimensions and their kinetic therms are canonically normalized in equation (\ref{eq:S4d}).  The potential of the scalar field is given by
\be
V(\tilde{\phi}^a) = n\langle E_1(\tilde{\phi}^a/M_p)\rangle \; ,
\ee
where $\langle E_1(\tilde{\phi}^a/M_p)\rangle$ can be approximated in the same limits as in Section \ref{sec:mod}.  For the purpose of our analysis, we will use the field redefinition $\tilde{\phi}^a = \tilde{\phi}_0^a + \Delta\tilde{\phi}^a$ where $\tilde{\phi}_0^a = (2M_p)^{-1}\ln \alpha'$ is the value of $\tilde{\phi}^a$ at the self-dual point.  With this field redefinition, the mass $M_{1,-1,1}^2$ can conveniently written as
\be
M_{1,-1,1}^2 = M_s^2\left(\sum_a e^{2\frac{\Delta\tilde{\phi}^a}{M_p}} + \sum_a e^{-2\frac{\Delta\tilde{\phi}^a}{M_p}} - 2p\right) \; .
\ee
We now use the massless approximation of section \ref{sec:modsg} ($\langle E_1 \rangle \approx M_{1,-1,1}$) to study the potential of our system and compute the quantity $|\partial_a V|/V$, which is of constrained by the de Sitter conjecture. When $|\Delta\phi^a/M_p| \ll 1$, the potential reaches zero exponentially and $|\nabla V|/V$ blows up to infinity.  In the opposite limit, the potential quickly becomes dominated by either the momentum or winding modes and can be approximated as
\be
V(\Delta \tilde{\phi}^a) = \sum_a nM_se^{\frac{|\Delta \tilde{\phi}^a|}{M_p}} \; .
\ee
The potential increases exponentially at a rate on the order of $M_p^{-1}$ no matter which direction you go in the moduli space away from the self-dual point. Therefore, far from the self-dual point, the potential obeys the de Sitter criterion
\be
\frac{|\partial_aV|}{V} \sim \frac{1}{M_p} \geq \frac{c}{M_p} \; .
\ee
What remains to do now is to derive the exact value of the constant $c$ in our case.  We know that $|\partial_aV|/V$ has the mirror symmetry $\Delta \tilde{\phi}^a = - \Delta \tilde{\phi}^a$ and asympotically reaches a constant of order $M_p^{-1}$ far from the self-dual point.  Therefore, let us assume $\Delta \tilde{\phi}^a \gg 0$.  In this case, $|\partial_aV|/V$ is written as
\be
\frac{|\partial_aV|}{V} = \frac{1}{M_p}\frac{\sqrt{\sum_a e^{\frac{2\Delta \tilde{\phi}^a}{M_p}}}}{\sum_a e^{\frac{\Delta \tilde{\phi}^a}{M_p}}} \; .
\label{eq:deSitter1}
\ee
It's easy to show that the gradient of the equation above is zero when all $\Delta \tilde{\phi}^a$'s are the same ($\Delta \tilde{\phi}^a = \Delta \tilde{\phi}^b$).  In this case, equation (\ref*{eq:deSitter1}) reads
\be
\frac{|\partial_aV|}{V} = \frac{1}{\sqrt{p}}\frac{1}{M_p} \; .
\ee
Note at away from the line $\Delta \tilde{\phi}^a = \Delta \tilde{\phi}^b$, $|\partial_aV|/V$ takes values which are very close to $1/M_p$ except near certain lines passing trough the self-dual point.  For example, when $\Delta \tilde{\phi}^a = \Delta \tilde{\phi}^b$ but one of the $\Delta \tilde{\phi}^a$'s is zero, $|\partial_aV|/V$ will reach a value close to $\frac{1}{\sqrt{p-1}}\frac{1}{M_p}$ and so on. Aside from these other local minima, the absolute minium remains $\frac{1}{\sqrt{p}}\frac{1}{M_p}$.

\section{Conclusion and Discussion}

We studied some aspects of moduli stabilization using string gases in the context of the Swampland.  When included as matter in the effective field theory context, the string gas yields as a potential that stabilizes the moduli (the scalar fields which emerge from string theory) at the self-dual radius of the extra dimensions.  As a result, gravity emerges as the weakest force (in agreement with the {\it weak gravity conjecture}) and the {\it de Sitter conjecture} on the slope of the scalar field potential is naturally satisfied. Our analysis yields the value of the numerical coefficient $c$ which arises in the de Sitter conjecture. Its value is $1 / \sqrt{p}$. \footnote{It is important to mention a caveat when comparing our results to the de Sitter conjecture (we thank C. Vafa for stressing this point): the de Sitter conjecture concerns properties of the bare potential. The potential we are studying, on the other hand, is an effective potential which assumes the presence of a gas of string modes with momenta and windings about the compactified dimensions. On the other hand, this potential plays the same role in stabilizing the string moduli as the potentials usually considered in string inflation models. Hence, it is fair to compare our potential to the ones which are usually studied.} 

It would be interesting to apply similar arguments to string gases using other compactifications (e.g. Calabi-Yau manifolds). Based on our qualitative arguments, we expect the main conclusions to be unchanged, but it would be interesting to study the value of the constant $c$ in the {\it de Sitter conjecture} bound which emerges. It would also be interesting to study the constraints on the potentials for other moduli fields. Based on the analysis in \cite{Brandenberger:2005bd}, we expect that similar conclusions can be derived for shape moduli fields.

Our analysis supports the view that it will be very difficult to embed cosmological inflation into string theory: the scalar field potentials which emerge are too steep to support inflation with sufficiently slow rolling. The challenge for primordial inflation has recently been strengthened by the {\it Trans-Planckian Censorship Conjecture} (TCC) \cite{Bedroya} which shows that \cite{Bedroya2} that a phase of inflation which can explain the origin of structure in the universe is only possible if the scale of inflation is smaller than about $10^{10} \, {\rm GeV}$. Demanding the correct power spectrum for cosmological perturbations leads to further constraints which, in particular, yield an upper bound on the tensor to scalar ratio $r$ of about $r < 10^{-30}$ \footnote{These conclusions can be mitigated by adding new whistles to the inflationary scenario \cite{whistles}.}. String theory may hence favor an alternative early universe scenario such as {\it string gas cosmology} \cite{BV}, a scenario which assumes an initial high-temperature state of strings in thermal equilibrium, and yields an alternative to cosmological inflation for producing a spectrum of nearly scale-invariant cosmological perturbations \cite{NBV} with a slight red tilt, and predicts a nearly scale-invariant spectrum of gravitational waves with a slight blue tilt \cite{BNPV}, a prediction with which the predictions of string gas cosmology can be distinguished from those of canonical inflation models (see e.g. \cite{SGrev} for a review).

\section*{Acknowledgements}

This research was supported in part by funds from NSERC and the Canada Research
Chair program. RB thanks C. Vafa for discussions and comments on the draft.  SL thanks Heliudson Bernardo for useful discussions, NSERC for a graduate scholarship, the MSI for a graduate award, and Eran Palti and Miguel Montero for useful lectures on the Swampland at SIFTS 2019.

\end{document}